\documentclass[letterpaper,twocolumn,10pt]{article}

\usepackage{usenix,epsfig,endnotes}
\usepackage{booktabs}
\usepackage{multirow}
\usepackage{algorithm}
\usepackage{listings}

\usepackage{amsmath, amsfonts, graphicx, url, hyperref, xspace}
\usepackage{algpseudocode}
\usepackage{caption}
\usepackage{subcaption}

\usepackage{pgfplotstable,filecontents}
\pgfplotsset{compat=1.9}

\usepackage{longtable}

\usepackage{color}

\definecolor{mygreen}{rgb}{0,0.6,0}
\definecolor{mygray}{rgb}{0.5,0.5,0.5}
\definecolor{mymauve}{rgb}{0.58,0,0.82}

\usepackage{xspace}

\renewcommand{\paragraph}[1]{\medskip\noindent{\bf{#1.}}}
\newcommand{\specialcell}[2][c]{%
\begin{tabular}[#1]{@{}c@{}}#2\end{tabular}}
\newcommand{\code}[1]{\texttt{#1}}

\begin{document}

\date{}

\title{\Large \bf Debloating Software through Piece-Wise Compilation and Loading}

\author{
{\rm Anh Quach}\\
Binghamton University\\
aquach1@binghamton.edu
\and
{\rm Aravind Prakash}\\
Binghamton University\\
aprakash@binghamton.edu
\and
{\rm Lok Yan}\\
Air Force Research Laboratory\\
lok.yan@us.af.mil
}

\maketitle

\thispagestyle{empty}

\pagenumbering{gobble}

\subsection*{Abstract}
Programs are bloated. Our study shows that only 5\% of  \code{libc} is used on average across the Ubuntu Desktop environment (2016 programs); the heaviest user, \code{vlc} media player, only needed 18\%.  

In this paper: (1) We present a debloating framework built on a compiler toolchain that can successfully debloat programs (shared/static libraries and executables). Our solution can successfully compile and load most libraries on Ubuntu Desktop 16.04. (2) We demonstrate the elimination of over 79\% of code from \code{coreutils} and 86\% of code from SPEC CPU 2006 benchmark programs without affecting functionality. We show that even complex programs such as \code{Firefox} and \code{curl} can be debloated {\em without a need to recompile}. (3) We demonstrate the security impact of debloating by eliminating over 71\% of reusable code gadgets from the \code{coreutils} suite, and show that unused code that contains {\em real-world vulnerabilities} can also be successfully eliminated without adverse effects on the program. (4) We incur a low load time overhead.

\section{Introduction}\label{sec:intro}

Reusing code is a common and indispensable practice in software development. 
Commonly, developers follow a “one-size-fits-all” methodology where features are packaged into reusable code modules (e.g., libraries) that are designed to service multiple diverse sets of clients (or applications). 
While this model aids the development process, it presents a detrimental impact on security and performance. 
A majority of clients may not use all of the functionalities. 
For example, the standard C library (\code{libc}) is intended to be widely useful, and usable across a broad spectrum of applications although not all features are used by all applications.  
Clients must bear the burden of carrying all the features in the code with no way to disable or remove those features.

This extraneous code may contain its own bugs and vulnerabilities and therefore broadens the overall attack surface. 
Additionally, these features add unnecessary burden on modern defenses (e.g., CFI) that do not distinguish between used and unused features in software. 

Accumulation of unnecessary code in a binary \--- either by design (e.g., shared libraries) or due to software development inefficiencies \--- amounts to code bloating.
As a typical example, shared libraries are designed to contain the union of all functionality required by its users. 

Static dead-code-elimination \--- a static analysis technique used to identify unused code paths and remove them from the final binary \--- employed during compilation is an effective means to reduce bloat. 
In fact, under higher levels of optimization, modern compilers (clang, gcc) aggressively optimize code to minimize footprint. 
However, a major limitation to static dead-code elimination is that dead code in dynamically linked libraries cannot be removed; shared libraries are pre-built and are not analyzed by the loader. Inter-module dependency information is not available either. 
As a result, a large fraction of overall bloat occurs in shared libraries.  
Alternatively, programs can be statically linked (to apply dead-code elimination), but there are two main hurdles: patches to libraries require recompilation of all programs, which is not feasible, and licenses such as (L)GPL can complicate redistribution. 
Dynamic linking is key to practical and backwards-compatible solutions. 

To exemplify the security impact of bloating, consider \code{libc}, a Swiss Army knife in the arsenal of an attacker~\cite{shacham:2007:turingrop}. Suppose we are to implement a minimal program that simply exits and does nothing else. In assembly, this program will only contain three instructions ({\tt mov \$1, \%eax; mov \$0, \%ebx; int \$0x80}). 
However, a gcc compiled program will require the entirety of \code{libc} ($>$165k instructions) despite the fact that only the entry point handler is needed.

This is true for any of the several flavors of \code{libc} such as \code{glibc} and \code{musl-libc}. 
If we were able to detect this case and remove the rest of the \code{libc} code, then CFI and other solutions would be more effective since there are fewer control flows to analyze. Reusable gadgets originating from unused code are automatically removed due to debloating and attack characteristics for detection can be refined and confined to the smaller code base and behavior space. All of this hinges on the ability to remove unused code.

In this paper, we introduce a generic inter-modular late-stage debloating framework. 
As a primary contribution, our solution combines static (i.e., compile-time) and dynamic (i.e., load-time) approaches to systematically detect and automatically eliminate unused code from program memory. 
We do this by removing unused and therefore unnecessary code (by up to 90\% in some test cases). This can be thought of as a runtime extension to {\em dead code elimination}.
As a direct impact, our solution significantly increases the effectiveness of current software defense by drastically reducing the amount of code they must analyze and protect. 

We identify and remove unused code by introducing a {\em piece-wise compiler} that not only compiles code modules (executables, shared and static objects), but also generates a dependency graph that retains all compiler knowledge on \textit{which} function depends on \textit{what other} function(s). 
Traditional loaders will simply ignore the section, but our {\em piece-wise loader} will read the dependency information and will only dynamically load functions that are needed by a program. 
The dependency information is written to an optional ELF section. 
Here, and in the rest of this paper, we use the generalized term ``code module" to signify a shared library, static library or an executable and ``loader" to signify both loader and dynamic linker. 

\paragraph{CFI vs Piece-wise}
Piece-wise compilation and loading is not a replacement for CFI. It is an orthogonal solution that reduces attack space by performing cross-module code reduction with zero runtime overhead.
This not only reduces the amount and diversity of available gadgets, but more importantly, it reduces the amount of code to be analyzed by other defenses and thus significantly amplifies their security impact.
For example, our study shows that on average only 5\% of \code{libc} functions are imported by a program. Therefore, in conjunction with piece-wise, CFI and other gadget removal defenses (e.g.~\cite{pappas:2012:smashgadgets}) only need to analyze 5\% of \code{libc} code. 
In essence, \code{libc} protected by both piece-wise and CFI exposes significantly less attack space than \code{libc} protected by only CFI.
Moreover, CFI primarily provides exploit mitigation and no post-compromise protection, whereas by eliminating unused code, piece-wise prevents execution of unused code even after compromise. This is why we believe piece-wise is complementary to CFI.

Our contributions:
\begin{enumerate}
	\setlength\itemsep{0em}
	\item We perform a comprehensive study of how \code{glibc} and other shared libraries are used in a set of over 2016 diverse programs across different domains (e.g., http server, database, MPEG players, document editors) in Ubuntu Desktop 16.04. A detailed and lateral study across multiple libraries can be found in our prior work~\cite{quach2017study}. We report that in the average case 95\% of code in \code{glibc} is never used. To the best of our knowledge, we are the first to conduct such a study for \code{glibc}.
	\item We implement an LLVM-based piece-wise compiler that retains dependency information and generates backwards-compatible ELF files. Our compiler handles inlined assembly and implements three different independent approaches to capture indirect code pointers. We also introduce a backward compatible piece-wise loader that eliminates bloat. 
	\item Applying our toolchain to GNU \code{coreutils}, we eliminiate over 79\% of code and 71\% of ROP gadgets in \code{musl-libc} while passing all the tests accompanied by the  \code{coreutils} suite. Our solution introduces a low load-time overhead.
	\item We demonstrate that several {\em real world} vulnerabilities in unused code can be successfully eliminated using our piece-wise compiler and loader. 
\end{enumerate}

The rest of this paper is organized as follows.
Section~\ref{sec:background} provides details and results from our study of shared library usage.
Section~\ref{sec:overview} gives an overview as well as challenges and design goals of our methodology for late-stage debloating.
Sections~\ref{sec:pwc} and~\ref{sec:pwl} describes in details each part of our toolchain in details.
We evaluate the piece-wise prototype in Section~\ref{sec:eval}. 
Finally, we discuss related works in section~\ref{sec:related} and conclude in section~\ref{sec:conclude}.

\section{Bloating}\label{sec:background}

\noindent
{\bf Study:}
Code bloating occurs when a program contains excess unused code in its address space.
To get a sense of how pervasive and serious bloating is, we conducted a study encompassing all the userspace programs in Ubuntu Desktop 16.04. For each program, we 1) identified all libraries the program depends on using {\tt ldd}; 2) identified all functions imported by the program and which library the symbol can be found in as well as their intra-modular dependencies; and 3) for each dependent library, identified the exported functions that were never imported by the program. In essence, we recursively traversed through all dependent code modules of a program and gathered all the function-level dependencies. 

On average, only 10.22\% of functions in the top 15 most used shared libraries are used by programs (full results in Appendix~\ref{sec:appendix}). 
In the case of the most utilized (i.e., least bloated) library \code{libstdc++}, only 37.77\% of the library is used. On the other extreme, as low as 4\% of code in \code{libgcc} is used. 
Furthermore, Table~\ref{tab:2243cf} contains a list of programs that best utilize \code{libc}, i.e., contained the largest footprint within \code{libc}. Even \code{vlc} player \--- the least bloated program in the study \--- only used 18\% of code loaded into memory.

\begin{table}[]
	\centering
	\caption{Code footprint in \code{libc} corresponding to a subset of programs in the study. The mean reflects the geometric mean of all programs in the study.}
	\label{tab:2243cf}
	\resizebox{\linewidth}{!}{%
		\begin{tabular}{|l|l|l|l|l|}
			\hline
			{\bf Program} & {\bf \# Functions} & {\bf \# Insns} & \specialcell{\bf \% Fn \\ \bf Footprint} & \specialcell{\bf \% Insn \\ \bf Footprint} \\
			\hline
			vlc & 606 & 33371 & 21\% & 18\% \\
			rhythmbox & 579 & 28517 & 20\% & 16\% \\
			unopkg.bin & 520 & 27576 & 19\% & 16\% \\
			gst-xmlinspect-0.10 & 542 & 30184 & 19\% & 17\% \\
			kubuntu-debug-installer & 531 & 29258 & 19\% & 16\% \\
			soffice.bin & 543 & 29723 & 19\% & 17\% \\
			checkbox-gui & 525 & 28044 & 19\% & 15\% \\
			VBoxTestOGL & 500 & 26219 & 18\% & 15\% \\
			ktrash & 492 & 25621 & 18\% & 14\% \\
			kchmviewer & 504 & 27530 & 18\% & 15\% \\
			kdebugdialog & 503 & 27468 & 18\% & 15\% \\
			kwalletd & 506 & 27557 & 18\% & 15\% \\
			nepomukmigrator & 503 & 27468 & 18\% & 15\% \\
			kdesu & 519 & 27822 & 18\% & 15\% \\
			signon-ui & 498 & 27074 & 18\% & 15\% \\
			spotydl & 510 & 26406 & 18\% & 14\% \\
			webapp-container & 513 & 26516 & 18\% & 15\% \\
			knetattach & 510 & 27598 & 18\% & 15\% \\
			nepomukbackup & 512 & 27637 & 18\% & 15\% \\
			notepadqq-bin & 504 & 27280 & 18\% & 15\% \\
			
			... & ... & ... & ... & ...  \\
			\hline\hline
			Mean &176 & 9904 & 6\% & 5\% \\
			\hline
		\end{tabular}
	}%
\end{table}

\subsection{Root Causes of Bloating}
We report four main causes of bloating that we discovered through our study.

\paragraph{Multiple Disjoint Functionalities} 
By design, code modules may pack multiple functionalities that may be disjoint.
For example, libc provides subroutines for memory management (e.g., {\tt malloc, calloc, free}), file I/O (e.g., {\tt fopen, fclose, printf, scanf}), string manipulation (e.g., {\tt strcpy, toupper, tolower}), etc. 
In fact, we found as many as 30 different disjoint features packaged within libc (see Appendix~\ref{sec:appendix}). 

\paragraph{Backwards Compatibility}
Modern toolchains support backwards compatibility through a technique called {\em weak aliasing}. 
A {\em weak alias} signifies to the loader that a particular function should be used only when a better implementation (strong alias) does not exist. 
If available, the dynamic linker will bind the symbol names to the strong definitions, rendering the weak definitions redundant; the unused weak implementation remains in memory and contributes to bloating.

For example, \code{glibc} 2.19 hosts 610 (29\%) functions that are marked as weak symbols including popular memory management functions like {\tt calloc}.
In our study, we found that complex software like \code{Firefox} and \code{mongodb} provide custom implementations for memory management functions and override the one provided in \code{glibc}. 
This situation manifests in all cases where a functionality in one code module has a stronger binding than code in another module. 

\paragraph{Static Function Clones}
In C/C++, the {\tt static} keyword is used to limit the scope of a function or variable within the file in which it is defined. Due to the nature of how the {\tt \#include} preprocessor directive works, whenever a static function is defined within a header file, the compiler generates a copy of the function for each include. 
Furthermore, since static functions are local to a file, they do not trigger compile-time name conflicts.

\paragraph{Unused Functions}
Static analysis during compilation can efficiently remove dead code at a basic block level, however, entire unused functions are not eliminated. 
Consider the following program:
\begin{verbatim}
int f() { return 1; }
int main() { return 0; }
\end{verbatim}
Both gcc and clang retain the function {\tt f} in the above code even under optimization level -O3.
Removal of unused functions require additional non-standard often-unused compiler ({\tt -fdata-sections -ffunction-sections -Os}) and linker ({\tt -Wl,--gc-sections}) optimization flags.
Even so, unused functions in dynamically loaded libraries can not be eliminated during compile time. 

\section{Overview}\label{sec:overview}
\subsection{Key Challenges}
Debloating requires precise identification of program-wide intra- and inter-modular dependencies, which introduces several challenges: 
\begin{enumerate}
	\setlength\itemsep{0em}
	\item {\bf Modular Interdependencies:} Programs can depend on one or more dynamically linked shared libraries and each shared library may depend on other shared libraries. In essence, the library level dependencies can be viewed as a directed graph with cycles. The actual code path or function level dependencies is similar to context-sensitive inter-procedural analysis, a known hard problem in program analysis.
	\item {\bf Late binding:} The binding between a function symbol and the actual library that provides the functionality is not known until run-time. Furthermore, function binding depends on load order and potential use of {\em weak} symbols.
	\item {\bf Code-pointer within libraries:}
	Typically, calls between shared libraries, or a shared library and the main executable are routed through the PLT. However, dependencies between functions within libraries may not be apparent if code pointers are used to invoke functions, especially if such invocations happen within hand-written assembly code. Similar to CFI, a practical solution must correctly detect and include {\em all} dependencies arising from code pointer accesses within shared libraries. 
	\item {\bf Dependencies within hand-written assembly code:} 
	Generating inter-dependencies for assembly code in a module at compile time is challenging because assembly code is not analyzed by the compiler, and function boundaries in optimized code are sometime slurred.
	\item {\bf Dynamically loaded libraries:} 
	Shared libraries can be dynamically loaded at runtime using \code{dlopen}. 
	The use of this feature causes incomplete dependency information at program load time, which in turn impacts correctness.
	We use a combination of static analysis and training-based approach to preload and debloat dynamically loaded libraries.
\end{enumerate} 

The techniques presented in this paper are common to all code modules (i.e., shared and statically linked libraries, and executables).
Yet, the impact of piece-wise compilation and loading is best realized in shared libraries.
This is because while existing compile- and link-time optimizations can eliminate unused code within a compilation unit, bloat arising due to dynamically loaded modules persists due to the vast amounts of disjoint functionalities in shared libraries. 

At first glance, dynamically linked libraries are designed for code reuse (e.g., one copy of a library is resident in memory for multiple processes) and fine-grained function-level fragmentation of libraries in which each function and its dependencies are encapsulated within a single shared library may be an appealing solution. 
For example, if a program uses only {\tt printf}, then the {\tt printf} library that only contains {\tt printf} and its dependencies will be loaded. 
However, like in the static case, this design is not ideal for usability since each focused shared library is likely to be much smaller than the usual 4k page size granularity. This will result in heavy internal fragmentation, and much of the memory will remain unused.
Moreover, with such a design, complex software is likely to require hundreds if not thousands of shared libraries. 
Consequently, load-time and runtime relocations are likely to be high. 
Also, such a solution is not backward compatible and the programs linked to use shared libraries will now have to be recompiled to use multiple smaller libraries. 

\subsection{High Level Approach}
At a high-level, our approach bridges the traditional information gap between early (compilation) and late (loading) stages of a program.
Specifically, (1) we develop a piece-wise compiler that maintains intra-modular (piece-wise) dependencies between each individual functionality (i.e., entry point) and all dependent functions that are necessary to satisfy execution, and (2) we develop a piece-wise loader that examines the dependencies of an executable and generates an inter-modular full-program dependency graph. Finally, the loader systematically eliminates all code that is not a part of the full-program dependency graph. 

Our approach maintains the benefits of dynamically linked libraries (e.g., code-reuse) with the benefits of statically built programs (e.g., dead-code elimination). It is driven by these high-level goals:

\paragraph{Program-Wide Dead Code Elimination}
Our first goal is to support load-time dead-code elimination. That is, we aim to bring dead-code elimination benefits of static linking to dynamic linking.
In our approach, we analyze and embed functionality-specific metadata into code modules during compilation. Specifically, the metadata contains functions and all of the dependencies that are required to be loaded together with it in order to provide correct program execution. 
At runtime, when a program or library requests a new symbol to be loaded, we use the metadata to only load the dependent functionality. Unused code (code that does not have a runtime dependency) is never available to the program.

\paragraph{Backwards Compatibility}
We wish to allow existing binaries to reap the benefits of load-time dead-code elimination by debloating the dependent shared libraries,  {\em without the explicit need to recompile the entire program}. 
To retain backwards compatibility, we embed the metadata into an optional section in the ELF file format. Optional sections are ignored by unmodified loader, meaning our ELF files are backwards compatible with older loaders. As one would expect, our piece-wise loader is able to make use of this extra information to achieve late-stage code removal during loading. 
This way, any COTS software can take advantage of our piece-wise technique by simply replacing the shared libraries in a system with piece-wise compiled shared libraries and replacing the loader with our piece-wise loader.

\paragraph{Correctness}
It is essential that the solution be conservative and retain \textit{all} fragments of code within each code module that the program may need during runtime. Missing legitimate code dependencies will cause unacceptable runtime program failures. We wish to prevent such failures.

\section{Piece-wise Compilation}\label{sec:pwc}
For a given code module, the piece-wise compiler has two main tasks: generate a function-level dependency graph with zero false negatives (we do not want to miss any legitimate dependency), and write this dependency graph to the binary. 
\subsection{Dependency Graph Generation} In traditional dead-code elimination, analysis is performed at the basic block level. Thus, a dependency graph is effectively an annotated inter-procedural control flow graph. 
This fine granularity is not necessary for our application since symbols are exported at a function granularity. Our dependency graph is therefore an annotated call graph. 

We use a two-step process to generate the dependency graph. 
First, we combine all object files and generate a single complete call graph for the entire module.
Then, we traverse the call graph to generate the dependencies for each exported function. 
Here, we leverage the inter-modular code analysis and optimization logic present in LLVM to derive function-level dependencies both within a compilation unit and across a module. Of particular importance is handling special cases that can affect the accuracy of the call graph. Below, we detail the treatment of such cases to ensure complete dependency recovery. 

Two factors can have a significant effect on the accuracy of a call graph: code pointers and jump tables, and hand-written assembly (this includes pure-assembly functions and inlined assembly).
Below, we provide details about each case as well as how we handle them.

\subsection{Handling Code Pointers/Indirect Branching} 
The piece-wise compiler uses the call graph analysis pass of LLVM to extract dependencies arising due to direct calls between functions. However, indirect code-pointer references require special handling. 
Like some CFI solutions, we take a conservative approach and include a set of all functions that could potentially be used as indirect branch targets. While one can assume that a function pointer can point to any valid function, this may not be necessary. To see why, we separate the problem into two cases - function pointers associated with symbols and those that are not associated with symbols.

Function pointers that target symbols can be directly identified as long as the target is internal to the module being compiled. That is, the module contains code that loads the target function address into the function pointer as a constant. In other words, while the pointer itself is not initialized until runtime, the target can be determined statically. 
Pointers that target external function (still associated with symbols) can be reconciled at load time when all of the external modules are loaded along with the symbol information. Our piece-wise compiler is designed to retain such information as well.

\begin{lstlisting}[caption={File IO in \code{musl-libc}},label=ex:code-musl]
struct _IO_FILE {
	...
	size_t (*write) (FILE *, char *, size_t);
};
static struct _IO_FILE f = {
	...
	.write = __stdout_write,
};
FILE *const stdout = &f;
static void close_file(FILE *f) {
	...
	if (f->wpos > f->wbase)
	f->write(f, 0, 0);
	...
}
\end{lstlisting}

Indirect code references can be classified into three categories. We handle all 3 categories: 
\begin{enumerate}
	\setlength\itemsep{0em}
	\item[{\bf C1}] {\em Reference to a function pointer}: In this category, a function address is assigned---either directly or through a function argument---to a variable by one instruction and is used later by another instruction (e.g., {\tt addr = \&foo; addr();}).
	\item[{\bf C2}] {\em Reference to a table of code pointers:}  Here, a table or an array of function pointers is addressed as a base+offset (e.g., {\tt void (*foo)[LEN]() = \&table; foo[4]();}). \\
	Jump tables, arrays of function pointers, and vtables in C++ are all examples of this category. 
	\item[{\bf C3}] {\em Reference to a composite structure:} A more complex case arises when code pointers are contained within structures. Consider the example in Listing~\ref{ex:code-musl}. Variable {\tt f} is a global \code{IO} structure that contains a pointer to the {\tt write} function. This variable is initialized as a global, but used in the {\tt close} function. References through composite structures are not uncommon, yet hard to detect.
\end{enumerate}

Additionally, function pointers are used to implement callback functions, and are passed as arguments during callback registrations (e.g., arguments to \texttt{signal}, \texttt{qsort}). 
Callbacks are also used to register initialization and termination functions of a process (e.g. \texttt{atexit}).
Pointers passed through function arguments reduce to {\bf C1} in inter-procedural analysis.
Function pointers are also used to implement subtype polymorphism of records. For example, in \code{libc}, a \code{`FILE'} struct with a set of function pointers is created for every IO operation. 

In order to obtain a complete set of code pointer references within a module, we perform code-pointer analysis (function pointer analysis + jump table recovery) to recover all potential code references either to functions or to code snippets (e.g., targets in switch statement).
We introduce two new independent approaches to handle indirect control-flow transfers: full-module code pointer scanning and localized code pointer scanning.
They are based on an observation that all functions serving as indirect targets must have their addresses taken at some point during execution. 
A function has its address taken when its address is referenced as a constant somewhere within a module. 
Additionally, we leverage well-studied points-to analysis techniques.
Comparison between these three approaches can be found in Section~\ref{sec:eval}.

\paragraph{Full-Module Code Pointer Scan}
In this approach, our compiler statically generates a global set of functions as global dependency for the entire module. 
Each instruction in the LLVM IR is scanned for code pointer references, and when a reference is found, the referenced code is recorded as a required global dependency.
The global dependency includes all functions that have their addresses referenced inside the module. These dependencies are annotated as ``required" in the optional section of the ELF binary, and therefore will be retained in memory at runtime. 
While this approach may not result in optimal code reduction, it is fast and is guaranteed to include all possible targets of indirect branches.

\paragraph{Localized Code Pointer Scan}
Similar to the full-module scan, the localized scan aims to include all possible indirect branch targets in the working module. 
However, we observe that among all code addresses that the compiler detects, only a selective few actually have their addresses taken at runtime; we can safely unload the rest of code pointers to boost debloating result, without loss of correctness. 
For example, suppose in the code snippet in Listing~\ref{lst:local-scan}, {\tt comp} is referenced {\em only} by function {\tt foo}. 
Then, {\tt comp} is marked as a dependency for {\tt foo}, and is retained if {\tt foo} is also retained.
Similarly, if multiple functions depend on {\tt comp}, it is added to the dependency graph of each function. 
This is unlike the full-module scan where {\tt comp} is marked as required for the entire module.
\begin{lstlisting}[caption={Localized Code Pointer Scan Example},label=lst:local-scan]
...
int comp(int a, int b) {...}
int foo() { ... /* foo is a global symbol */
sort(arr, len, &comp); }
...
\end{lstlisting}

First, {\em use-def} chains are constructed for all IR instructions. 
Here, unlike traditional use-def analysis, we are only interested in the referring nodes that directly take a function's address.
To accurately recover all instructions that use function address, our compiler recursively traverse the use-def chains until it encounters a referring-instruction that refers a function. 
At that point, a dependency is recorded between the function that contains the referring instruction and the referred function.
When compared to the full-module scan, by leveraging symbol binding information available, this approach improves dependency graph's correctness and debloats more aggressively, but at the cost of analysis performance.

\paragraph{Pointer Analysis}
We leverage points-to information produced by pointer analysis to resolve indirect code pointer dependencies within a library. 
Broadly, our approach is based on the inclusion-based algorithm first introduced by Andersen~\cite{andersen1994pta}, where a points-to set is maintained for each pointer variable. When an assignment {\tt a = b} is encountered, locations pointed to by {\tt b} are assumed to be a subset of locations pointed to by Our implementation is based on the algorithm recently proposed by Sui et al.~\cite{sui2016svf}. 
Each LLVM IR statement with a pointer reference is analyzed to extract rules that define how to generate points-to information. These form the constraints. 
We extract four types of constraints that were first proposed by Hardekopf and Lin~\cite{hardekopf2009semi} based on semantics of the pointer reference.
For convenience, we include a reproduction in Table~\ref{tab:constraints} below. 
	\begin{table}[ht]
		\centering
		\caption{Points-to constraints. For a variable $v$, $pts(v)$ represents $v$'s points-to set and $loc(v)$ represents the memory location denoted by $v$.}
		\label{tab:constraints}
		\resizebox{\linewidth}{!}{%
			\begin{tabular}{| c | c | c | }
				\hline
				{\bf Program Code} & {\bf Constraint} & {\bf Meaning} \\ \hline\hline
				$a$ = \&$b$ & a $\supseteq$ \{$b$\} & $loc(b) \in pts(a)$ \\ 
				$a = b$ & $a \supseteq b$ & $pts(a) \supseteq pts(b)$ \\
				$a = *b$ & $a \supseteq *b$ & $\forall v \in pts(b) : pts(a) \supseteq pts(v)$ \\
				$*a = b$ & $*a \supseteq b$ & $\forall v \in pts(a) : pts(v) \supseteq pts(b)$ \\
				\hline
			\end{tabular}
		}%
	\end{table}
These constraints are then fed into a constraint solver to extract concrete pointer values/value sets at different code-pointer reference points within functions.
These pointers form dependencies for the functions. 
We refer readers to SVF~\cite{sui2016svf} for additional details.

\paragraph{Object-Sensitive Analysis for C++ Code}
Due to virtual function dispatch in C++, indirect code pointers that are referenced through a VTable require special handling. 
Two separate solutions are considered.
First, a naive solution would be to include (and persist in memory) all functions in all VTables. While such an approach will include all required dependencies, it fails to provide optimal bloat reduction. 

For the second approach, we introduce object-sensitive analysis in Algorithm~\ref{alg:cpp} to identify precise virtual function dependencies.
\begin{algorithm}[ht]
	\begin{small}
		\caption{Gathering virtual function dependencies in C++ code. Function $GetFunctionDeps$ recursively traverses call graph to provide a complete list of dependencies for a given function.} \label{alg:cpp}
		\begin{algorithmic}[1]
			\Procedure{GetDependencies}{$Function$}
			\State $Deps \gets \emptyset$
			\For{{\bf each} $DepFunc \in GetFunctionDeps(Function)$}
			\State $Deps \gets Deps \cup GetDependencies(DepFunc)$
			\EndFor
			\For{{\bf each} $Object \in Function$} \Comment Function instantiates Object
			\State $VTable \gets GetVTable(TypeOf(Object$))
			\For{{\bf each} $VFunc \in VTable$}
			\State $Deps \gets Deps \cup VFunc$
			\EndFor
			\EndFor
			\State \Return $Deps$
			\EndProcedure
		\end{algorithmic}
	\end{small}
\end{algorithm}
For each function within the dependency graph, we examine the code to identify all the types of C++ objects that are instantiated within the function and gather the corresponding VTables. 
Next, for each type of object, we include all of the virtual functions in the VTable for the corresponding class as a dependency for the function that instantiates the object. 
This way, if an object is never instantiated, its VTable functions are debloated.
Finally, we incorporate in our solution pointer analysis to handle C++ virtual dispatch.

\subsection{Handling Assembly Code}
Compilers do not optimize hand-written and inline assembly code and, as such, interdependencies involving assembly code are handled separately.  

\vspace{.08in}
\noindent
{\em Dependencies in assembly code:}
We perform a single pass through assembly code to identify all function calls and update the callgraph accordingly.
From our experiments, we find that this simple approach is sufficient to capture all the higher-level (e.g., C/C++) function dependencies for code originating from assembly. 

\vspace{.08in}
\noindent
{\em Dependencies on assembly code:}
Identifying assembly code dependencies for high-level functions is more difficult since function boundaries in optimized code is sometimes blurred due to code reuse. 
For example, some functions jump directly into the middle of the assembly code for {\tt memcpy} instead of calling {\tt memcpy} directly. 
We take a conservative approach and retain all assembly code as necessary.
As such, assembly code is never removed from memory. 
Handwritten assembly is uncommon and therefore including it does not significantly impact bloat.

\subsection{Writing Dependency Graph to Binary}
Once the dependency graph is generated, it is embedded into a dedicated section called \texttt{.dep}.
Our compiler inserts two types of information to assist the loader with identifying dead code: dependency relationships between functions (i.e. the dependency graph) that comprises of functions and a list of dependencies, and function-specific data that includes location and size in bytes for all the functions in the dependency graph.
Since a function's address is unknown at link time, we instead mark all location fields in \texttt{.dep} section as relative relocatable and let the loader patch them with real addresses during program load time. 
While the piece-wise compiler only embeds function dependency information in binary, it can retains more information to assist precise late-stage security enforcement such as CFI. 

\section{Piece-Wise Loader}\label{sec:pwl}
\begin{figure*}[ht]
    \centering
    \includegraphics[width=.8\linewidth]{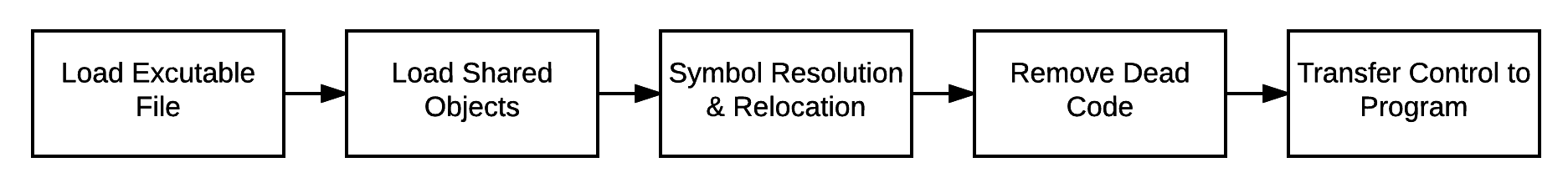}
    \caption{Workflow of the piece-wise loader}
    \label{fig:loader}
\end{figure*}

Figure \ref{fig:loader} illustrates the workflow of our piece-wise loader. After receiving control from the kernel, the loader first maps all dependent libraries onto the current process' address space, then performs relocation on all modules, and finally eliminates all dead code from piece-wise-compiled libraries. Our current implementation readily supports position independent code and can be easily deployed in current Linux ecosystems.

\subsection{Pre-Loading Dependencies} 
In order to generate a complete set of all exported library functions that a program requires, the piece-wise loader must resolve the dependencies within the program executable along with all the other shared objects the executable depends on. 
Since loaders are designed to load libraries when they are first used, some libraries may not be loaded when the program starts. This results in incomplete symbol information. To address this, our loader pre-loads all shared libraries.

First, the piece-wise loader recursively traverses all shared objects and their dependencies (by looking at \code{DT\_NEEDED} entries of the dynamic section of the ELF file of the program executable) to construct the list of shared objects that the main program needs. Then, it maps their memory segments onto the process image. Effectively, a program and all of its dependent code are loaded into memory {\em before} transferring control to the user code.

\paragraph{Handling Dynamically Loaded Libraries}
Dynamically loaded libraries create function dependencies that are unknown during both compile time (and therefore are not encoded in dependency graphs) and load time. 
Thus, as a result of late-stage piece-wise debloating, such functions are removed and unavailable in cases where dynamically loaded libraries
require them.
Support for shared libraries that are loaded dynamically (using {\tt dlopen}) proves to be a challenge. 
On the one hand, for cases where we can statically detect which libraries will be dynamically loaded, i.e. arguments to functions like \code{dlopen} are hard-coded in binaries, we directly pre-load them.
On the other hand, handling dynamically generated library names is challenging. An example of such case can be found in Listing~\ref{lst:dlopen}:

\begin{lstlisting}[caption={Example of dynamically generated library name.},label=lst:dlopen]
lib_name = compute_lib_name();
handle = dlopen(lib_name, RTLD_NOW);
\end{lstlisting}

Failure to accommodate for the library's dependencies will cause a runtime failure. 
However, the non-determinism makes ensuring absolute correctness intractable.
Therefore, we take a training-based approach to identify all missing dependency caused by dynamic loading. 
For each program, we record all shared libraries loaded using \code{dlopen} at runtime as well as their functions that are invoked by \code{dlsym} and embed this information within the binaries.
At load time, the piece-wise loader will interpret it, pre-load those libraries, and retain only the functions that \code{dlsym} invokes.

We found that only 64/2226 (2.9\%) programs in our study dynamically compute module names. In our test set, all library name computations are straightforward: library names are hard-coded or generated using format string. For example, \code{if (var) sprintf(name, "lib\%s\_v1.so", basename) else sprintf(name, "lib\%s\_v2.so", basename)}. In our experience, training for common workloads reveal required shared-lib dependencies.

\subsection{Symbol Resolution \& Relocation} 
After loading the libraries and performing the necessary symbol bindings, the loader walks through the dependency information in the {\tt .dep} section and marks code as necessary. 
All unnecessary code is zeroed out. 
Recall that the dependency information in the optional {\tt .dep} section contains the symbol as well as its location in the binary and size. 
In order to support relocation of the piece-wise compiled libraries, these locations must be updated prior to resolving all dependencies. 
Handling relocation for \code{.dep} section is straightforward. 
Traditionally, at load time, the loader will walk through all relocatable fields in a mapped ELF image and patch them with appropriate addresses.
We simply ensure that the same procedure also applies to the optional {\tt .dep} section and updates its relocatable fields.

Recall that loaders prioritize the resolution of strong symbols over weak ones. Therefore, if two libraries offer bindings to the same symbol, the first strong symbol is resolved --- this depends on the order of which shared libraries are loaded. As a result, the behavior is also runtime dependent. 

Since we pre-load libraries in the order they appear in an ELF file, symbol resolution is also performed in the same order.
This process, called pre-binding, ensures that each required symbol is bound to the concrete definition in the executable or a shared library {\em before} the program begins execution.
Therefore, all dependencies for a program are known before it begins execution.

To determine which functions are not required at runtime, i.e., the ones that must be removed, we rely on symbol resolution and the dependency graph embedded in the \code{.dep} section. 
During symbol resolution, the loader binds an undefined symbol to the first available definition for the symbol in the load order which allows our loader to identify which library functions the program imports.

At the end of symbol resolution, all symbols in the global symbol table are fully resolved and reflect the runtime necessities of the program. 
If there are two different definitions of the same symbol name in two separate code modules, only one will be picked; we can safely zero out the other.
For example, if {\tt foo.exe} depends on function {\tt myFoo}, which is defined in both shared libraries \code{a.so} and \code{b.so}, the symbol is resolved to whichever library is loaded first.
That is, if \code{a.so} is loaded before \code{b.so}, then {\tt myFoo} in \code{b.so} is never used, and is therefore removed.
The dependency graph in the \code{.dep} section for each resolved symbol is used to determine precisely which further dependencies to retain.
For example, if {\tt myFoo} is resolved to \code{a.so}, and {\tt myFoo}'s dependency contains function {\tt myBar} in \code{a.so}, then {\tt myBar} will be retained alongside {\tt myFoo} in \code{a.so}.

The result generated from this step is a list of functions to be removed from each library.

\subsection{Removal of Dead Code}
There are two approaches to eliminating dead code: either we start with a clean canvas and load each required function and its dependencies, or we load the entire module and remove dead code. To support shared libraries, since most code and data references are relative due to position-independent code, we implement the latter in our prototype. 
This preserves the offset between functions and therefore does not require any unnecessary code modifications. 

All functions in a piece-wise module that do not form direct or indirect dependencies are marked for removal. 
If all the code in a page is marked for removal, we simply set the non-executable bit on the page and no code deletion is performed. 
To remove a certain function, the loader invokes {\tt mprotect} to mark the corresponding code page(s) as writable and non-executable. 
Next, every byte in the function body is set to a special 1-byte invalid instruction. 
In the x86 and x86\_64 architectures, we pick byte 0x6d since it is a reserved instruction that raises an {\em `Illegal Instruction'} exception. 
Once all unused functions are removed, a piece-wise library is rendered bloat-free.

\paragraph{Backward compatibility}
Both piece-wise-compiled modules and the piece-wise loader are backward compatible for two reasons.
First, our changes are restricted to the optional {\tt .dep} section in a code module while all other sections remain intact. 
Therefore, a regular loader simply ignores the {\tt .dep} section and skips support for debloating. 
Second, when the piece-wise loader loads a code module without the {\tt .dep} section, it simply behaves like a regular non-piece-wise loader. 
No modifications are required to the program being executed as long as the program is configured to use the piece-wise loader.
This can be accomplished by patching the {\tt .interp} section of the ELF binary and changing it from the default loader (e.g. \code{/lib/ld-linux.so}) to the pathname of the new piece-wise loader {e.g. \code{/lib/pw-linux.so}}. 

\paragraph{Memory overhead due to copy-on-write}
When the piece-wise loader marks an entire page as non-executable, it incurs no memory overhead. 
An overhead (due to CoW) is incurred when partial removal occurs in a page. 
Because large fractions of code are typically eliminated from the memory, very few pages actually require CoW. 
In general problems arise when "multiple" long-lived processes share large libraries, or when unused code is distributed across multiple pages. 
While we did not engineer the support for dynamically rewriting the binary to reduce memory overhead, we refer interested readers to artificial diversity research for an algorithm~\cite{larsen2014diversity}.

\section{Evaluation}\label{sec:eval}
We divide our evaluation into three main parts: debloating correctness (sections \ref{subsec:debloatlibc} and \ref{subsec:debloatcurl}), performance overhead (section \ref{subsec:perf}), and impact of debloating on security (section \ref{subsec:cfi}). 
Because our solution neither adds executable code in the program nor alters the code layout, we do not introduce any runtime execution overhead. 
All of our experiments were performed on a system with Intel Core i7-4790 @ 3.60GHz and 32GB RAM running Ubuntu Desktop 16.04 LTS.

\begin{table*}[ht]
	\footnotesize
	\centering
	\caption{Percentage Attack Space Reduction with Piece-Wise for  \code{coreutils} and SPEC CPU 2006 with \code{musl-libc}.}
	\label{tab:dyn-code-reduction}
	\begin{tabular}{|l|ll|ll|ll|}
		\hline
		\multirow{2}{*}{\bf Program} & \multicolumn{2}{l|}{\bf Full-module Code Pointer Scan} & \multicolumn{2}{l|}{\bf Inclusion-based Pointer Analysis} & \multicolumn{2}{l|}{\bf Localized Code Pointer Scan} \\ \cline{2-7} 
		& \specialcell{\bf \% Function \\ \bf Reduction} & \specialcell{\bf\% Instruction \\ \bf Reduction} & \specialcell{\bf \% Function \\ \bf Reduction} & \specialcell{\bf \% Instruction \\ \bf Reduction} & \specialcell{\bf \% Function \\ \bf Reduction} & \specialcell{\bf \% Instruction \\ \bf Reduction} \\ \hline
		\specialcell{Minimal Program} & 60 & 60 & 89 & 91 & 88 & 91 \\
		\specialcell{Coreutils Min} & 59 & 59 & 85 & 85 & 84 & 85 \\
		\specialcell{Coreutils Max} & 60 & 60 & 88 & 90 & 88 & 91 \\
		\specialcell{Coreutils Mean} & 56 & 58 & 79 & 78 & 79 & 79 \\
		\hline
		bzip2 & 60 & 60 & 89 & 90 & 88 & 91 \\
		sjeng & 59 & 59 & 85 & 86 & 85 & 86 \\
		sphinx3 & 59 & 60 & 86 & 85 & 81 & 82 \\
		mcf & 60 & 60 & 85 & 83 & 87 & 87 \\
		lbm & 58 & 59 & 83 & 83 & 87 & 87 \\
		gcc & 60 & 60 & 87 & 87 & 84 & 87 \\
		milc & 59 & 59 & 88 & 88 & 84 & 85 \\
		h264ref & 60 & 60 & 88 & 87 & 84 & 83 \\
		hmmer & 60 & 60 & 85 & 85 & 82 & 83 \\
		gobmk & 60 & 60 & 86 & 86 & 85 & 86 \\
		libquantum & 58 & 58 & 81 & 82 & 87 & 89 \\ \hline
		\specialcell{SPEC CPU 2006 \\ Mean} &  59 & 60 & 86 & 86 & 85 & 86 \\
		\hline
	\end{tabular}
\end{table*}

\subsection{Implementation and Prototype}
We implemented two different versions of piece-wise loaders: (1) the GNU loader (v2.23)  distributed with Ubuntu Desktop 16.04, and (2) the loader packaged within \code{musl-libc} (v1.1.15). 
Because \code{glibc} can not be compiled using LLVM, we used \code{musl-libc} for the C library debloating evaluation. 
Accordingly, the GNU loader was used in experiments where \code{glibc} was used (the modified loader debloated libraries other than \code{glibc}), and the \code{musl} loader was used to debloat programs that used \code{musl-libc}. 
Both loaders were designed to retain and load non-piece-wise compiled libraries without any changes. 

The piece-wise compiler is built on top of LLVM-4.0 with an additional 2.46 KLOC.
First, we added an LLVM module pass to handle code pointers, process points-to information (if applicable), parse function calls from assembly code and generate a dependency graph.
Second, to support C++ libraries, we implemented an object-sensitive approach described in Algorithm~\ref{alg:cpp}.
We evaluated our C++ libraries debloating on \code{libflac++} using \code{Audacity}, a program editing audio files.
Our analysis and dependency graph generation and insertion passes are run during the link-time optimization (LTO) in LLVM gold plugin. 
We also developed an ELF binary patching program that patches an ELF binary to modify the {\tt .interp} section to change the default loader to the piece-wise loader. 

\subsection{Correctness Experiments}
To demonstrate that our toolchain correctly debloats code modules, we used the piece-wise compiler to build 400 shared libraries distributed with Ubuntu Desktop 16.04 and installed them using \code{dpkg}.
Next, we replaced the GNU loader with our piece-wise loader.

Below, we consider each set of libraries to gain a better understanding of the effectiveness and security benefits that our solution offers. 

\begin{table*}[ht]
	\footnotesize
	\centering
	\caption{Gadget reduction in  \code{coreutils} 8.2 and SPEC CPU 2006 benchmarks for 6 different types of security sensitive gadgets: syscall, stack pointer update (SPU), call-oriented programming (COP), call-site/call preceded gadgets(CS), jump-oriented programming (JOP), and entry-point (EP). For each type, we list the quantity found in debloated musl-libc and the percentage reduction achieved by piece-wise toolchain. In vanilla musl-libc, we found a total of 5619 unique gadgets, 485 syscall, 924 SPU, 334 COP, 780 CS, 47 JOP, and 22 EP. }
	\label{fig:gadget}
	\resizebox{\linewidth}{!}{%
	\begin{tabular}{|l|ll|ll|ll|ll|ll|ll|ll|}
		\hline
		\textbf{Program} & \multicolumn{2}{l|}{\bf Total} & \multicolumn{2}{l|}{\bf syscall} & \multicolumn{2}{l|}{\bf SPU} & \multicolumn{2}{l|}{\bf COP} & \multicolumn{2}{l|}{\bf CS} & \multicolumn{2}{l|}{\bf JOP} & \multicolumn{2}{l|}{\bf EP} \\ \hline
		Minimal Program & 993 & 82.33\% & 106 & 78.14\% & 147 & 84.09\% & 80 & 76.05\% & 109 & 86.03\% & 18 & 61.70\% & 4 & 81.82\% \\ 
		\hline
		\hline
		coreutilts max & 1971 & 64.92\% & 205 & 57.73\% & 325 & 64.83\% & 182 & 45.51\% & 253 & 67.56\% & 24 & 48.94\% & 5 & 77.27\% \\
		 \code{coreutils} min & 1274 & 77.33\% & 117 & 75.88\% & 187 & 79.76\% & 119 & 64.37\% & 149 & 80.90\% & 21 & 55.32\% & 4 & 81.82\% \\
		 \code{coreutils} mean & 1591 & 71.69\% & 142 & 70.75\% & 245 & 73.45\% & 138 & 58.67\% & 186 & 76.15\% & 23 & 51.02\% & 4 & 81.60\% \\ 
		\hline
		\hline
		bzip2 & 1256 & 77.65\% & 108 & 77.73\% & 185 & 79.98\% & 111 & 66.77\% & 150 & 80.77\% & 21 & 55.32\% & 4 & 81.82\% \\
		gcc & 1749 & 68.87\% & 144 & 70.31\% & 285 & 69.16\% & 156 & 53.29\% & 210 & 73.08\% & 26 & 44.68\% & 4 & 81.82\% \\
		gobmk & 1545 & 72.50\% & 141 & 70.93\% & 246 & 73.38\% & 137 & 58.98\% & 177 & 77.31\% & 21 & 55.32\% & 4 & 81.82\% \\
		h264ref & 1467 & 73.89\% & 120 & 75.26\% & 220 & 76.19\% & 130 & 61.08\% & 165 & 78.85\% & 21 & 55.32\% & 4 & 81.82\% \\
		hmmer & 1499 & 73.32\% & 130 & 73.20\% & 230 & 75.11\% & 133 & 60.18\% & 173 & 77.82\% & 24 & 48.94\% & 4 & 81.82\% \\
		lbm & 1685 & 70.01\% & 125 & 74.23\% & 259 & 71.97\% & 183 & 45.21\% & 204 & 73.85\% & 26 & 44.68\% & 4 & 81.82\% \\
		libquantum & 1570 & 72.06\% & 125 & 74.23\% & 239 & 74.13\% & 144 & 56.89\% & 174 & 77.69\% & 23 & 51.06\% & 4 & 81.82\% \\
		mcf & 1367 & 75.67\% & 119 & 75.46\% & 203 & 78.03\% & 128 & 61.68\% & 159 & 79.62\% & 21 & 55.32\% & 4 & 81.82\% \\
		milc & 1810 & 67.79\% & 166 & 65.77\% & 274 & 70.35\% & 199 & 40.42\% & 243 & 68.85\% & 25 & 46.81\% & 4 & 81.82\% \\
		sjeng & 1417 & 74.78\% & 122 & 74.85\% & 202 & 78.14\% & 133 & 60.18\% & 165 & 78.85\% & 21 & 55.32\% & 4 & 81.82\% \\
		sphinx3 & 1398 & 75.12\% & 120 & 75.26\% & 199 & 78.46\% & 127 & 61.98\% & 161 & 79.36\% & 21 & 55.32\% & 4 & 81.82\% \\
		\hline
		\specialcell{SPEC CPU 2006 \\ Mean} & 1,524 & 72.88\% & 129 & 73.38\% & 231 & 74.99\% & 144 & 56.97\% & 180 & 76.91\% & 23 & 51.64\% & 4 & 81.82\% \\ \hline
	\end{tabular}
}%
\end{table*}

\subsubsection{Musl-libc Experiements}\label{subsec:debloatlibc}
Due to known fundamental limitations in compiling \code{glibc} using LLVM\cite{glibc_clang}, we piece-wise compiled \code{musl-libc}---another popular and comprehensive flavor of  the C library.
The difference in functionality between \code{glibc} and \code{musl-libc} does not affect the feasibility and capability of the piece-wise toolchain.

To get a sense of how much \code{glibc} can be debloated, we extracted 30 different features and the functions within each feature from the \code{glibc} software development manual~\cite{libc-doc}, and mapped them to analogous symbols in \code{musl-libc}.
We piece-wise compiled \code{musl-libc}, and computed the footprint for each category. 
Our findings are tabulated in Table~\ref{tab:bf-categories} and a corresponding cumulative distribution is represented in Figure~\ref{fig:cfp} in Appendix~\ref{sec:appendix}.

The virtual memory allocation and paging related functions are most widely used, but only account for 1.91\% of instructions. 
Similarly, string related functions are second most widely used, but contribute only 5.82\% of instructions. 
This result solidifies our findings from the pervasiveness study in Section~\ref{sec:background}, and highlights the vast amounts of unused \code{libc} code in typical program memory.
Mathematics (different from Arithmetic) contributes the most code, but is seldom used. 
We expect \code{glibc} to be just as bloated due to the functional similarities between \code{glibc} and \code{musl-libc}.
Unfortunately, due to constraints in building \code{glibc}~\cite{glibc_clang} we are unable to provide concrete evidence at this time.

\begin{table*}[]
	\footnotesize
	\centering
	\caption{Percentage Attack Space Reduction for 14 piece-wise libraries used by \code{curl} program. }
	\label{tab:curl-red}
	\resizebox{\linewidth}{!}{%
		\begin{tabular}{@{}|l|ll|ll|ll|@{}}
			\hline
			\multirow{2}{*}{\bf Library} & \multicolumn{2}{l|}{\bf Full-module Code Pointer Scan} & \multicolumn{2}{l|}{\bf Inclusion-based Pointer Analysis} & \multicolumn{2}{l|}{\bf Localized Code Pointer Scan} \\ \cline{2-7}
			& \specialcell{\bf \% Function \\ \bf Reduction} & \specialcell{\bf \% Instruction \\ \bf Reduction} & \specialcell{\bf \% Function \\ \bf Reduction} & \specialcell{\bf \% Instruction \\ \bf Reduction} & \specialcell{\bf \% Function \\ \bf Reduction} & \specialcell{\bf \% Instruction \\ \bf Reduction} \\ \hline
			libasn1 & 21.15\% & 41.85\% & 22.01\% & 42.18\% & 22.01\% & 42.17\% \\
			libcurl & 3.43\% & 2.30\% & 28.57\% & 40.79\% & 25.14\% & 39.74\% \\
			libgssapi & 7.70\% & 9.67\% & 14.96\% & 26.11\% & 38.62\% & 73.12\% \\
			libheimbase & 7.37\% & 9.15\% & 11.54\% & 21.38\% & 25.64\% & 50.86\% \\
			libheimntlm & 14.06\% & 34.45\% & 14.06\% & 34.46\% & 14.06\% & 34.45\% \\
			libheimsqlite & 0.63\% & 0.17\% & 2.68\% & 1.59\% & 17.23\% & 11.30\% \\
			libhx509 & 18.39\% & 35.25\% & 24.40\% & 44.40\% & 35.89\% & 65.05\% \\
			libidn & 19.84\% & 20.77\% & 19.84\% & 20.77\% & 19.84\% & 20.77\% \\
			libkrb5 & 13.98\% & 18.49\% & 21.55\% & 30.45\% & 26.73\% & 41.44\% \\
			libp11-kit & 7.14\% & 11.07\% & 63.07\% & 74.95\% & 58.21\% & 65.78\% \\
			librtmp & 21.05\% & 21.50\% & 21.05\% & 21.51\% & 22.22\% & 22.30\% \\
			libtasn1 & 16.76\% & 31.34\% & 16.76\% & 31.35\% & 16.76\% & 31.34\% \\
			libwind & 8.75\% & 16.23\% & 15.00\% & 19.95\% & 8.75\% & 16.23\% \\
			libz & 35.61\% & 35.97\% & 35.61\% & 36.15\% & 37.07\% & 43.21\% \\
			\hline
			\hline
			Mean & 13.99\% & 20.59\% & 22.22\% & 31.86\% & 26.30\% & 39.84\% \\
			\hline
		\end{tabular}
	}
\end{table*}

\paragraph{Debloating coreutils}
Using the piece-wise compiled \code{musl-libc}, we tested \code{coreutils} to evaluate correctness and performance. 
All of the programs (109 in total) in  \code{coreutils} passed the  \code{coreutils} test suite that is packaged with
 \code{coreutils} source code without errors. 
Table \ref{tab:dyn-code-reduction} shows the percentage of attack space reduction achieved with piece-wise on  \code{coreutils} programs and a minimal program for each code pointer handling approach. 
The minimal program contains a main function that immediately returns. 
Percentage of attack space reduction achieved with minimal program serves as a lower bound for debloating \code{musl-libc}. 
Our results show that, among the three approaches for handling code pointers, localized code pointer scan and pointer analysis achieve the best debloating result (79\% and 78\% respectively) while full-module debloats the least, 58\%.
For some programs, (e.g., \code{make-prime-list}), 91\% of \code{libc} code was removed without errors for localized scan.

\paragraph{Debloating SPEC CPU2006 benchmark programs}
Similarly, in order to verify correctness, we also evaluated SPEC CPU2006 benchmark programs using piece-wise compiled \code{musl-libc} with all three code pointer handling approaches. Results are tabulated in Table~\ref{tab:dyn-code-reduction}. 
We note that the latest version of \code{musl-libc} does not fully support the SPEC CPU2017 benchmarks.
All of the programs ran successfully and passed the reference workload. In the best case, 86\% attack space reduction was achieved with localized scan and pointer analysis, and in the worst case, 60\% code reduction was achieved for full-module pointer scan. 

While on average, pointer analysis and localized code pointer scan yield the same attack space reduction results, for some cases in the SPEC CPU 2006 benchmarks, we observe that one outperformed the other.
Because localized code pointer scan records the relationships between the functions that contains referencing instructions and the referenced functions, the piece-wise loader will only remove an address taken function if all referring functions are removed.
Thus, this approach takes advantage of symbol resolution information only available at program load time.
On the one hand, the localized scan approach provides better debloating results when it allows removing functions that will not have address taken at runtime because all referring functions have been removed while pointer analysis does not.
On the other hand, pointer analysis debloats more than localized scan when the number of retained address taken functions is larger than the size of points-to set.

\subsubsection{Debloating COTS binaries}\label{subsec:debloatcurl}
In order to demonstrate the efficacy of our approach on COTS binaries, we debloated unmodified programs in the Ubuntu 16.04 Desktop environment. 
First, we piece-wise compiled a set of shared libraries (minus \code{glibc}). Then, we replaced the default loader with the piece-wise loader, and the default libraries with the piece-wise compiled libraries. 
A subset of the shared libraries with various compile-time overheads are presented in Table~\ref{tab:llvm-pass-time}.

First, we confirmed that the piece-wise loader was able to successfully load unmodified shared libraries.
Next, we manually tested a variety of unmodified executables --- \code{FireFox}, \code{curl}, \code{git}, \code{ssh} and \code{LibreOffice} programs that used the piece-wise compiled libraries. We were able to verify that the loader correctly loaded the piece-wise compiled libraries, and all of them ran under normal use without errors.
The bloat reduction results for \code{curl} are tabulated in Table~\ref{tab:curl-red} for each code pointer handling approach.
Despite not debloating \code{glibc}, we were able to reduce bloat by over 39.84\% on average for localized scan. 
In general, libraries that are general purpose are more bloated (e.g., \code{libasn1}) than the libraries that are a part of the application pacakge (e.g., \code{libcurl}).
We demonstrate that a COTS binary which uses \code{glibc} can still be debloated, even if \code{glibc} is not piece-wise compiled. 
We show that our solution can target some if not all shared libraries used by a program, and is truly backward compatible.

\subsubsection{Debloating C++ Libraries}
To demonstrate piece-wise seamless support for C++ code, we successfully compiled and debloated \code{libFLAC++}.
We were able to successfully remove 46.09\% of functions or 66.90\% of instructions. 
Debloating results are summarized in table~\ref{tab:c++lib}.

\begin{table*}[ht]
\footnotesize
\centering
\caption{Debloating \code{libFLAC++} with \code{Audacity}.}
\label{tab:c++lib}
\begin{tabular}{|l|ll|ll|ll|}
\hline
\specialcell{\bf Handling Technique} & \specialcell{\bf \# Removed \\ \bf Functions} & \specialcell{\bf \# Removed \\ \bf Instructions} & \specialcell{\bf \# Functions \\ \bf Total} & \specialcell{\bf \# Instructions \\ \bf Total} & \specialcell{\bf \% Function \\ \bf Reduction} & \specialcell{\bf \% Instruction \\ \bf  Reduction} \\ \hline
\hline
\specialcell{Object-sensitive, \\ Inclusion-based \\Pointer Analysis} & 271 & 5831 & 588 & 8716 & 46.09\% & 66.90\% \\ \hline
\end{tabular}
\end{table*}

\begin{table*}[ht]
	\centering
	\caption{Vulnerabilities Removed after Debloating Libraries}
	\label{tab:vuln}
	\resizebox{\linewidth}{!}{%
	\begin{tabular}{|l|l|l|l|l|}
		\hline
		\bf{Library} & \bf{CVE-ID} & \bf{Functions Affected} & \bf{Program} & \bf{Vulnerability Type} \\ \hline
	
		zlib-1.2.8 & CVE-2016-9842 & inflateMark & \specialcell{git, curl, \\ LibreOffice, firefox} & Undefined Behavior \\ \hline
		\multirow{3}{*}{libcurl-7.35} & CVE-2016-7167 & \specialcell{curl\_escape, \\ curl\_easy\_escape, \\ curl\_unescape, \\ and curl\_easy\_unescape} & curl & Integer Overflow \\ \cline{2-5} 
		& CVE-2014-3707 & curl\_easy\_duphandle & curl, cmake & Out-of-bound Read, Use After Free \\ \cline{2-5} 
		& CVE-2016-9586 & curl\_mprintf & cmake & Buffer Overflow \\ \hline
	\end{tabular}
}%
\end{table*}

\subsubsection{Piece-wise vs Static Linking}
While static linking provides optimal debloating benefits, its use in practice is limited due to the following reasons:
\begin{itemize}
	\setlength\itemsep{0em}
	\item Requires recompilation of binaries with every library or software update.
	\item Does not allow memory sharing across processes.
	\item May result in accidental violation of (L)GPL.
	\item Increases binary size compared with dynamic linking.
	\item Risks transferring bugs in a shared library to the binary.
\end{itemize}
Since piece-wise aims to bring dead code elimination benefits from static linking to dynamic linking, in table~\ref{tab:pwisevsstatic}, we compare whole-program code reduction achieved by static linking with late-stage debloating using piece-wise toolchain.
The percentage reduction in this table takes into account both program and library code to accurately delineate program-wise debloating of both approaches.
Static linking provides an upper bound for dead code elimination.
Localized code pointer scan was able to remove most of the code from program's address space, followed by pointer analysis and full-module scan. 
Overall, we observe that piece-wise's dead code elimination benefit is comparable but not as efficient as static linking due to analysis accuracy and the retention of necessary code for piece-wise loading and code removal.

\begin{table*}[]
	\centering
	\caption{Whole-process attack space reduction of static linking and piece-wise for  \code{coreutils} and SPEC CPU 2006.}
	\label{tab:pwisevsstatic}
	\begin{tabular}{|l|l|l|l|l|}
		\hline
		\textbf{Program} & \textbf{Static Linking} & \textbf{Pointer Analysis} & \textbf{Localized Scan} & \textbf{Full-module Scan} \\ \hline
		Minimum Program & 99.55\% & 95.67\% & 96.11\% & 63.42\% \\ \hline
		coreutils mean & 81.42\% & 76.18\% & 78.22\% & 54.97\% \\ \hline
		bzip2 & 84.28\% & 78.38\% & 81.18\% & 43.33\% \\
		gcc & 14.13\% & 13.10\% & 13.57\% & 7.55\% \\
		gobmk & 39.37\% & 36.55\% & 37.84\% & 21.28\% \\
		h264ref & 44.94\% & 41.78\% & 43.28\% & 25.06\% \\
		hmmer & 59.05\% & 55.13\% & 57.00\% & 33.13\% \\
		lbm & 88.75\% & 82.33\% & 85.24\% & 47.16\% \\
		libquantum & 87.23\% & 80.86\% & 83.77\% & 45.61\% \\
		mcf & 89.66\% & 83.24\% & 86.15\% & 47.66\% \\
		milc & 75.26\% & 70.05\% & 72.49\% & 41.25\% \\
		sjeng & 76.76\% & 71.39\% & 73.89\% & 41.36\% \\
		sphinx3 & 68.72\% & 64.50\% & 66.47\% & 38.82\% \\ \hline
	\end{tabular}
\end{table*}

\subsection{Performance Overhead}\label{subsec:perf}
\paragraph{Compile-time overhead}
We measured execution time added by our LLVM pass for each of the three approaches (full-module scan, localized scan and inclusion-based points-to analysis) by inserting timing code at the beginning and end of pass' main logic.
The results are tabulated in Table~\ref{tab:llvm-pass-time}.
Full-module scan is the quickest followed by localized scan.
Both incur reasonable overhead (worst case $< 800ms$). Due to constraint-solving, points-to analysis was the slowest. 
In general, we found greater-than-linear increase in overhead introduced by points-to analysis with respect to the code size, with up to 4 minutes for \code{libheimsqlite.so}. 
While this is indeed a large overhead, we believe that this one-time overhead is reasonable given the large attack space reduction it provides (see Section~\ref{subsec:cfi}).

\begin{table}[ht]
	\centering
	\caption{Piece-wise LLVM Pass Execution Time. All entries are in milliseconds.}
	\label{tab:llvm-pass-time}
	\resizebox{\linewidth}{!}{%
		\begin{tabular}{|l|l|l|l|}
			\hline
			{\bf Library} & \specialcell{\bf Full-Module \\ \bf Code Pointer \\ \bf Scan} & \specialcell{\bf Inclusion-based \\ \bf Analysis} & \specialcell{\bf Localized \\ \bf Code Pointer \\ \bf Scan} \\ \hline
			musl-libc & 73 & 28661 & 158 \\
			libasn1 & 40.80 & 16,000 & 41.40 \\
			libcurl & 23 & 891 & 79.10 \\
			libgssapi & 14.10 & 31,600 & 132 \\
			libheimbase & 6.30 & 1,570 & 8.94 \\
			libheimntlm & 0.81 & 275 & 1.02 \\
			libheimsqlite & 406 & 241,000 & 3,380 \\
			libhx509 & 22.20 & 12,700 & 4.07 \\
			libidn & 0.67 & 0.68 & 0.68 \\
			libkrb5 & 165 & 20,700 & 776 \\
			libp11-kit & 6.95 & 4,330 & 0.89 \\
			librtmp & 2.66 & 1,000 & 3.31 \\
			libtasn1 & 2.19 & 1,370 & 2.36 \\
			libwind & 0.27 & 186 & 0.25 \\
			libz & 1.20 & 1,530 & 7.63 \\ \hline
		\end{tabular}
	}%
\end{table}

\paragraph{Load-time overhead}
Our changes to the loader, which eventually removes unused shared library code before transferring control to {\tt \_\_libc\_start\_main} only affects a program's start-up time. We do not add any code to the program's execution. 
Load time overhead caused by debloating comes from two sources. 
First, since we have added code to piece-wise loader to perform debloating, this extra logic introduces overhead to a program's load time.
To measure this, we ran each program in  \code{coreutils} sequentially, measured load time for default and piece-wise loaders, then computed the overhead.
On average, the code piece-wise loader that performs debloating added 20 milliseconds to the each process load time across all  \code{coreutils} programs.

Second, because piece-wise loader writes to code pages that contain the copies of shared libraries, copy-on-write is triggered, which results in additional load time overhead.
To measure debloating's effect on system with a large number of debloated processes running concurrently, we launched a number of programs in  \code{coreutils} simultaneously and measured the overhead caused by the piece-wise loader.
With all 106 programs running concurrently, we observed an overhead of 49 milliseconds for each process.
We are currently working on a solution to minimize the loadtime overhead. 

\subsection{Attack Space Reduction}\label{subsec:cfi}
\paragraph{Gadget Elimination}
While gadget reduction does not stop all attacks, it does give an estimate of how much attack space is reduced.
In Table~\ref{fig:gadget}, we show overall gadget reduction as well as reduction security-sensitive gadgets that have been extensively used in previously published work such as syscall~\cite{shacham:2007:turingrop}, stack-pointer update (SPU)~\cite{sotirov2007heap,Goktas:2014:CFI:Attack}, call-oriented programming (COP)~\cite{carlini:2014:ROP}, call-site/call preceded (CS)~\cite{Goktas:2014:CFI:Attack,carlini:2014:ROP} , jump-oriented programming (JOP)~\cite{bletsch:2011:jop}, and entry-point (EP)~\cite{Goktas:2014:CFI:Attack} gadgets.
This reduction is measured in \code{musl-libc} for \code{coreutils} and SPEC CPU 2006 benchmarks using ROPgadget~\cite{salwan:rop}. 
Overall, we were able to remove 71\% of gadgets.
Although we did not test for exploitation, elimination of high-impact gadgets will, in principle, hamper return-to-libc and code-reuse exploits. 

\paragraph{Vulnerability Elimination}
Another observable security benefit of removing unused code is that we also eliminate its vulnerabilities.
We perform an extensive study on all shared libraries we tested, analyzed all removed functions, and cross-referenced them with the list of reported CVE for each libraries.
Results are listed in table~\ref{tab:vuln}.

\subsection{Case Study: CVE-2014-3707}
\code{Curl} is a widely used program with known critical security vulnerabilities. In fact, over 25 vulnerabilities in \code{curl} have been reported in 2016 alone~\cite{curlvul}.
Similarly, the \code{curl} library used by many programs for handling file transfers (e.g. cmake, LibreOffice, git, Luau, and OpenOffice) has reported several vulnerabilities.
Our solution significantly reduces attack space through \code{libcurl} debloating and therefore offers several security benefits, one of which is vulnerability elimination as listed in table~\ref{tab:vuln}.
To demonstrate this, we show how an attacker can leak information using a vulnerability discovered in \code{libcurl} and how our solution defeats this through debloating.

CVE-2014-3707~\cite{curlcve3707} is an out-of-bound read vulnerability in function \code{curl\_easy\_duphandle} affecting \code{libcurl} versions 7.17.1 to 7.38.0 that can be exploited for memory disclosure and denial-of-service attacks.
\code{curl\_easy\_duphandle} uses \code{strdup} to copy buffers under the assumption that they are C strings terminated by \code{NULL}.
If the assumption is violated, \code{strdup} will read beyond buffers' boundaries, allowing an attacker to crash the program by triggering a segmentation fault or, in the worst case scenario, perform an out-of-bound memory read.
To make matters worse, after duplication, it fails to update the pointer to point to the new buffer which can trigger illegal use of freed memory if original object has been freed.

Our evaluation shows that debloating \code{libcurl} when it is used with programs like \code{curl} or \code{cmake} completely removes the affected functions and therefore the bug can no longer be exploited to perform a memory disclosure or a denial-of-service attack as part of an exploit payload such as through a return-to-libc attack.
We emphasize that our solution will not only eliminate known vulnerabilities but will also potentially remove yet-to-be-discovered ones.
This is one of the many security advantages that come with code debloating.

\section{Related Work}\label{sec:related}
\paragraph{Attack-Space Reduction Approaches}
Numerous efforts have attempted to defeat attacks by enforcing various forms of program properties such as SPI~\cite{prakash2015defeating,quach2017spiglass} and CFI as it decreases the size of the CFG and retains compile-time information.
CFI solutions extract the CFG and add instrumentation checks to the binary either by relying on source code and debugging information~\cite{Abadi:2005:CFI,vtvgcc:2012:tice}, or by analyzing the binary itself~\cite{Zhang:2013:CFI:Bin,Lenx:2013:CFI:Bin}.
Variations of CFI targeting either performance~\cite{qiao2015principled,Bletsch:2011:CFI,Zhang:2016:Vtrust}, or security~\cite{Mashtizadeh:2015:CCE:2810103.2813676,van2015practical} have been proposed.

ASLR~\cite{team2003pax,bhatkar:2003:aslr} was introduced as a means of preventing attackers from reusing exploit code effectively against multiple instantiations of a single vulnerable program. 
Wartell et al.~\cite{wartell:2012:stir} introduced binary stirring, which increases ASLR's re-randomization frequency to each time a program is launched.
Qiao et al.~\cite{qiao2015principled} interpret the ability to return to a location as a one-time capability, which is issued in each calling context in order to enable a one-time return. 
Niu and Tan~\cite{niu:2014:modular,niu:2014:rockjit,niu2015per} created a toolchain supporting fine-grained, per-input CFG generation and enforcement that combines dynamic linkin$  $g, support for JIT compilers and interoperability with unprotected legacy binaries. 
Giuffrida et al.~\cite{180231} presented a live re-randomization strategy for operating system load-time address space randomization to defend against return-into-kernel-text ROP attacks.
Crane et al.~\cite{7163059,crane2015s} uses a combination of compiler transformations and hardware-based enforcement to mark pages as execute-only, thereby defeating the objective of memory disclosures. 
Techniques that combine CFI and ASLR have also been proposed~\cite{mohan:2015:opaque}.
Piece-wise compilation and loading is independent of, yet complements CFI-based approaches.

\vspace{0.01in}
\paragraph{Feature-based Software Customization}
Unlike C/C++, managed programing languages whose execution is monitored by Runtime Virtual Machine suffers from significant runtime overhead or bloating due to the extra logic added to manage an execution environment. This bloating is categorized into two groups: memory bloat and execution bloat. Xu et al.~\cite{xu2010software} and Bu et al.~\cite{bu2013bloat} delegate the debloating task to developers, classifying this problem as purely software engineering related. On the other hand, Jiang et al.~\cite{jiang2015preliminary} propose a feature-based solution that allows a developer to remove certain feature in Java bytecode by performing static analysis. Jiang et al.~\cite{jiang2016jred} introduces an automatic approach that statically analyzes and removes unused codes in both Java application and Java Runtime Environment.
As a key distinction, our approach involves load-time dead-code removal to debloat shared libraries and reduce attack space in COTS binaries.

\paragraph{Pointer Analysis} Pointer analysis or points-to analysis, a well-studied and active research area, refers to determining memory targets of a pointer at compile time. Although precise flow-sensitive pointer analysis allows for high-quality and aggressive optimization, it is a proven NP-hard \cite{horwitz1997precise}. Numerous approaches have been proposed to balance the trade-off between performance/scalability and precision. A pointer analysis algorithm is classified based on various dimensions such as flow-sensitivity, context-sensitivity, intra/inter-procedural, and heap modeling. Flow-sensitive algorithms (\cite{hardekopf2011fspta}, \cite{yu2010fspta}, \cite{oh2012fspta}) take into account the control flow of a procedure; thus, the points-to information is more precise and different for each program point. However, a flow-insensitive points-to analysis (e.g. \cite{andersen1994pta} for inclusion-based and \cite{steensgaard1996points} for unification-based), is universal and refers to any execution points within a module. Similarly, context-sensitive analysis (e.g. \cite{whaley2004cspta}, \cite{wilson1995cspta}, \cite{xu2008cspta}) generates more precise points-to information by investigating each call site's context.

\section{Conclusion}\label{sec:conclude}
We presented a study across 2016 real world programs on Ubuntu Desktop 16.04 and show that most of the code in \code{libc} is seldom used. We implemented a prototype system that performs piece-wise compilation and loading. We evaluated the system and showed that \code{libc} can be debloated to eliminate significant code fragments from memory thereby reducing the attack space. 

\section{Acknowledgement}
We would like to thank the anonymous reviewers for their feedback.
This research was supported in part by Office of Naval Research Grant \#N00014-17-1-2929. Any opinions, findings and conclusions in this paper are those of the authors and do not necessarily reflect the views of the Office of Naval Research Grant and US government. 

\appendix
\section{Appendix}\label{sec:appendix}
Library-wise functional dependency is presented in Table~\ref{tab:study-libs}.
\begin{table}[ht]
	\centering
	\footnotesize	
	\caption{Most frequently used shared libraries in the study and their function-level code utility.}
	\label{tab:study-libs}
	\begin{tabular}{|l|c|c|}
		\hline
		{\bf Library} & {\bf \specialcell{\# programs\\that use the lib}} & {\bf Avg. \% of functions used} \\ \hline
		libc & 1932 & 24.64 \\
		libm & 284 & 7.06 \\
		libstdc++ & 266 & 37.77 \\
		libpthread & 237 & 11.10 \\
		libnetpbm & 201 & 4.74 \\
		libresolv & 186 & 9.60 \\
		libglib & 178 & 4.25 \\
		libtinfo & 170 & 12.42 \\
		libgio & 135 & 5.74 \\
		libdl & 125 & 4.18 \\
		libz & 116 & 6.07 \\
		libgcc & 113 & 4.0 \\
		libX11 & 89 & 6.04 \\
		libXau &  86 & 7.13 \\
		libselinux & 72 & 8.57 \\ \hline \hline
		& {\bf Mean (top 15):} & {\bf 10.22} \\ \hline		
	\end{tabular}
\end{table}

Functionality-size code footprint in \code{musl} is presented in Table~\ref{tab:bf-categories}.
\begin{table*}[ht]
	\centering
	\small
	\caption{Code footprint in \code{musl libc} by functionality categories.}
	\label{tab:bf-categories}
	\scalebox{0.7}{%
		\begin{tabular}{|l|ll|ll|}
			\hline
			Functionality & \# Functions & Function Footprint & \# Instructions & \specialcell{Instruction \\ Footprint} \\
			\hline
			Virtual Memory Allocation And Paging     & 14           & 1.45\%             & 1131            & 1.91\%                \\
			String and Array Utilities               & 74           & 7.65\%             & 3436            & 5.82\%                \\
			Mathematics                              & 175          & 18.10\%            & 20331           & 34.42\%               \\
			The Basic Program/System Interface       & 16           & 1.65\%             & 991             & 1.68\%                \\
			Pattern Matching                         & 103          & 10.65\%            & 7807            & 13.22\%               \\
			Date and Time                            & 29           & 3.00\%             & 1363            & 2.31\%                \\
			POSIX Threads                            & 4            & 0.41\%             & 104             & 0.18\%                \\
			Character Handling                       & 34           & 3.52\%             & 757             & 1.28\%                \\
			File System Interface                    & 49           & 5.07\%             & 1706            & 2.89\%                \\
			Low-Level Input/Output                   & 34           & 3.52\%             & 1783            & 3.02\%                \\
			System Configuration Parameters          & 4            & 0.41\%             & 242             & 0.41\%                \\
			System Management                        & 14           & 1.45\%             & 403             & 0.68\%                \\
			DES Encryption and Password Handling     & 5            & 0.52\%             & 354             & 0.60\%                \\
			Searching and Sorting                    & 15           & 1.55\%             & 686             & 1.16\%                \\
			Users and Groups                         & 40           & 4.14\%             & 890             & 1.51\%                \\
			Processes                                & 15           & 1.55\%             & 810             & 1.37\%                \\
			Resource Usage And Limitation            & 23           & 2.38\%             & 545             & 0.92\%                \\
			Job Control                              & 10           & 1.03\%             & 152             & 0.26\%                \\
			Inter-Process Communication              & 13           & 1.34\%             & 497             & 0.84\%                \\
			System Databases and Name Service Switch & 0            & 0.00\%             & 0               & 0.00\%                \\
			Non-Local Exits                          & 3            & 0.31\%             & 34              & 0.06\%                \\
			Message Translation                      & 12           & 1.24\%             & 741             & 1.25\%                \\
			Signal Handling                          & 17           & 1.76\%             & 448             & 0.76\%                \\
			Arithmetic Functions                     & 147          & 15.20\%            & 6974            & 11.81\%               \\
			Locales and Internationalization         & 4            & 0.41\%             & 208             & 0.35\%                \\
			Low-Level Terminal Interface             & 20           & 2.07\%             & 617             & 1.04\%                \\
			Syslog                                   & 5            & 0.52\%             & 196             & 0.33\%                \\
			Pipes and FIFOs                          & 4            & 0.41\%             & 263             & 0.45\%                \\
			Character Set Handling                   & 18           & 1.86\%             & 2724            & 4.61\%                \\
			Internal probes                          & 2            & 0.21\%             & 26              & 0.04\%                \\
			Sockets                                  & 52           & 5.38\%             & 2318            & 3.92\%                \\
			Error Reporting                          & 12           & 1.24\%             & 533             & 0.90\%            \\
			\hline   
		\end{tabular}
	}
\end{table*}

\begin{figure*}[ht]
	\centering
	\includegraphics[width=1\linewidth]{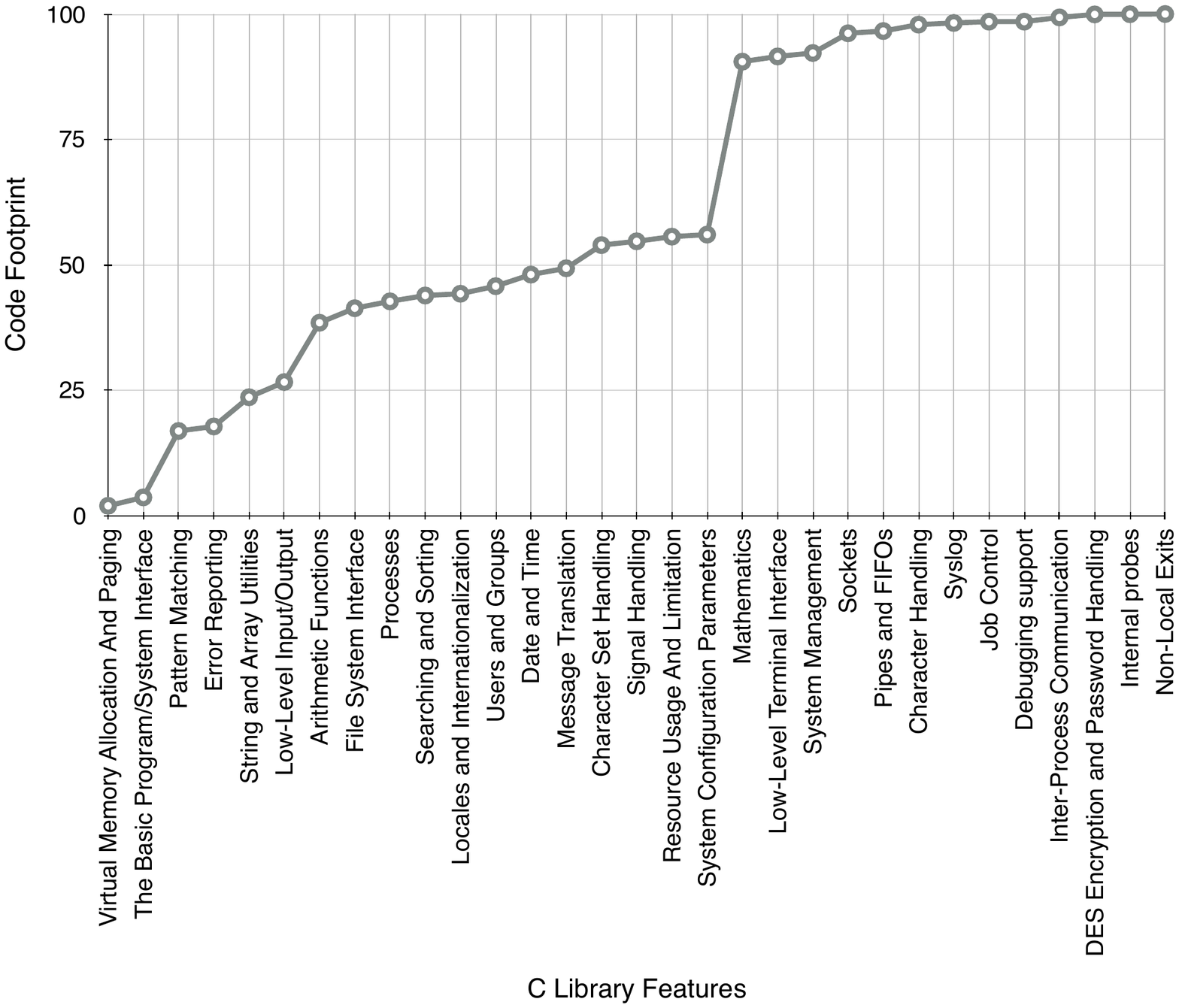}
	\caption{A cumulative distribution of code footprint in \code{libc} versus frequently used \code{libc} functions in our study. Virtual memory allocation and paging functionality is most used and non-local exits are least used. Figure shows all 30 features in \code{libc} }
	\label{fig:cfp}
\end{figure*}

\code{Musl} code footprint by features is presented in Figure~\ref{fig:cfp}.


\begin{thebibliography}{10}

\bibitem{glibc_clang}
Compiling glibc with clang/llvm.
\newblock https://groups.google.com/forum/\#\!topic/llvm-dev/arwzyPtQ2yY.
\newblock Accessed: 2017-01-20.

\bibitem{curlvul}
Curl: All known prior vulnerabilities.
\newblock https://curl.haxx.se/docs/security.html.

\bibitem{curlcve3707}
Cve-2014-3707.
\newblock https://cve.mitre.org/cgi-bin/cvename.cgi?name=CVE-2014-3707.

\bibitem{libc-doc}
The gnu c library: Categories and functions.

\bibitem{Abadi:2005:CFI}
{\sc Abadi, M., Budiu, M., Erlingsson, U., and Ligatti, J.}
\newblock {Control-flow Integrity}.
\newblock In {\em Proceedings of the 12th ACM Conference on Computer and
  Communications Security (CCS'05)\/} (2005), pp.~340--353.

\bibitem{andersen1994pta}
{\sc Andersen, L.~O.}
\newblock {\em Program analysis and specialization for the C programming
  language}.
\newblock PhD thesis, 1994.

\bibitem{bhatkar:2003:aslr}
{\sc Bhatkar, S., DuVarney, D.~C., and Sekar, R.}
\newblock {Address Obfuscation: An Efficient Approach to Combat a Broad Range
  of Memory Error Exploits}.
\newblock In {\em USENIX Security\/} (2003), vol.~3, pp.~105--120.

\bibitem{Bletsch:2011:CFI}
{\sc Bletsch, T., Jiang, X., and Freeh, V.}
\newblock Mitigating code-reuse attacks with control-flow locking.
\newblock In {\em Proceedings of the 27th Annual Computer Security Applications
  Conference\/} (2011), ACSAC '11, pp.~353--362.

\bibitem{bletsch:2011:jop}
{\sc Bletsch, T., Jiang, X., Freeh, V.~W., and Liang, Z.}
\newblock {Jump-Oriented Programming: A New Class of Code-Reuse Attack}.
\newblock In {\em Proceedings of the 6th ACM Symposium on Information, Computer
  and Communications Security\/} (2011), ACM, pp.~30--40.

\bibitem{bu2013bloat}
{\sc Bu, Y., Borkar, V., Xu, G., and Carey, M.~J.}
\newblock A bloat-aware design for big data applications.
\newblock In {\em ACM SIGPLAN Notices\/} (2013), vol.~48, ACM, pp.~119--130.

\bibitem{carlini:2014:ROP}
{\sc Carlini, N., and Wagner, D.}
\newblock {ROP} is still dangerous: Breaking modern defenses.
\newblock In {\em 23rd USENIX Security Symposium (USENIX Security'14)\/}
  (2014).

\bibitem{7163059}
{\sc Crane, S., Liebchen, C., Homescu, A., Davi, L., Larsen, P., Sadeghi,
  A.~R., Brunthaler, S., and Franz, M.}
\newblock Readactor: Practical code randomization resilient to memory
  disclosure.
\newblock In {\em 2015 IEEE Symposium on Security and Privacy\/} (May 2015),
  pp.~763--780.

\bibitem{crane2015s}
{\sc Crane, S.~J., Volckaert, S., Schuster, F., Liebchen, C., Larsen, P., Davi,
  L., Sadeghi, A.-R., Holz, T., De~Sutter, B., and Franz, M.}
\newblock It's a trap: Table randomization and protection against
  function-reuse attacks.
\newblock In {\em Proceedings of the 22nd ACM SIGSAC Conference on Computer and
  Communications Security\/} (2015), ACM, pp.~243--255.

\bibitem{180231}
{\sc Giuffrida, C., Kuijsten, A., and Tanenbaum, A.~S.}
\newblock Enhanced operating system security through efficient and fine-grained
  address space randomization.
\newblock In {\em Presented as part of the 21st USENIX Security Symposium
  (USENIX Security 12)\/} (Bellevue, WA, 2012), USENIX, pp.~475--490.

\bibitem{Goktas:2014:CFI:Attack}
{\sc G\"{o}kta\c{s}, E., Anthanasopoulos, E., Bos, H., and Portokalidis, G.}
\newblock Out of control: Overcoming control-flow integrity.
\newblock In {\em Proceedings of 35th IEEE Symposium on Security and Privacy
  (Oakland'14)\/} (2014).

\bibitem{hardekopf2009semi}
{\sc Hardekopf, B., and Lin, C.}
\newblock Semi-sparse flow-sensitive pointer analysis.
\newblock In {\em ACM SIGPLAN Notices\/} (2009), vol.~44, ACM, pp.~226--238.

\bibitem{hardekopf2011fspta}
{\sc Hardekopf, B., and Lin, C.}
\newblock Flow-sensitive pointer analysis for millions of lines of code.
\newblock In {\em Proceedings of the 9th Annual IEEE/ACM International
  Symposium on Code Generation and Optimization\/} (Washington, DC, USA, 2011),
  CGO '11, IEEE Computer Society, pp.~289--298.

\bibitem{horwitz1997precise}
{\sc Horwitz, S.}
\newblock Precise flow-insensitive may-alias analysis is np-hard.
\newblock {\em ACM Transactions on Programming Languages and Systems (TOPLAS)
  19}, 1 (1997), 1--6.

\bibitem{jiang2016jred}
{\sc Jiang, Y., Wu, D., and Liu, P.}
\newblock Jred: Program customization and bloatware mitigation based on static
  analysis.
\newblock In {\em Computer Software and Applications Conference (COMPSAC), 2016
  IEEE 40th Annual\/} (2016), vol.~1, IEEE, pp.~12--21.

\bibitem{jiang2015preliminary}
{\sc Jiang, Y., Zhang, C., Wu, D., and Liu, P.}
\newblock A preliminary analysis and case study of feature-based software
  customization.
\newblock In {\em Software Quality, Reliability and Security-Companion (QRS-C),
  2015 IEEE International Conference on\/} (2015), IEEE, pp.~184--185.

\bibitem{larsen2014diversity}
{\sc Larsen, P., Homescu, A., Brunthaler, S., and Franz, M.}
\newblock Sok: Automated software diversity.
\newblock In {\em 2014 IEEE Symposium on Security and Privacy\/} (May 2014),
  pp.~276--291.

\bibitem{Mashtizadeh:2015:CCE:2810103.2813676}
{\sc Mashtizadeh, A.~J., Bittau, A., Boneh, D., and Mazi\`{e}res, D.}
\newblock Ccfi: Cryptographically enforced control flow integrity.
\newblock In {\em Proceedings of the 22Nd ACM SIGSAC Conference on Computer and
  Communications Security\/} (New York, NY, USA, 2015), CCS '15, ACM,
  pp.~941--951.

\bibitem{mohan:2015:opaque}
{\sc Mohan, V., Larsen, P., Brunthaler, S., Hamlen, K., and Franz, M.}
\newblock {Opaque control-flow integrity}.
\newblock In {\em Symposium on Network and Distributed System Security
  (NDSS)\/} (2015).

\bibitem{niu:2014:modular}
{\sc Niu, B., and Tan, G.}
\newblock Modular control-flow integrity.
\newblock In {\em Proceedings of the 35th ACM SIGPLAN Conference on Programming
  Language Design and Implementation (PLDI'14)\/} (2014).

\bibitem{niu:2014:rockjit}
{\sc Niu, B., and Tan, G.}
\newblock Rock{JIT}: Securing just-in-time compilation using modular
  control-flow integrity.
\newblock In {\em Proceedings of 21st ACM Conference on Computer and
  Communication Security (CCS '14)\/} (2014).

\bibitem{niu2015per}
{\sc Niu, B., and Tan, G.}
\newblock Per-input control-flow integrity.
\newblock In {\em Proceedings of the 22nd ACM SIGSAC Conference on Computer and
  Communications Security\/} (2015), ACM, pp.~914--926.

\bibitem{oh2012fspta}
{\sc Oh, H., Heo, K., Lee, W., Lee, W., and Yi, K.}
\newblock Design and implementation of sparse global analyses for c-like
  languages.
\newblock {\em SIGPLAN Not. 47}, 6 (June 2012), 229--238.

\bibitem{pappas:2012:smashgadgets}
{\sc Pappas, V., Polychronakis, M., and Keromytis, A.~D.}
\newblock {Smashing the Gadgets: Hindering Return-Oriented Programming using
  in-place Code Randomization}.
\newblock In {\em IEEE Symposium on Security and Privacy (SP'2012)\/} (2012),
  pp.~601--615.

\bibitem{prakash2015defeating}
{\sc Prakash, A., and Yin, H.}
\newblock Defeating rop through denial of stack pivot.
\newblock In {\em Proceedings of the 31st Annual Computer Security Applications
  Conference\/} (2015), ACM, pp.~111--120.

\bibitem{qiao2015principled}
{\sc Qiao, R., Zhang, M., and Sekar, R.}
\newblock A principled approach for rop defense.
\newblock In {\em Proceedings of the 31st Annual Computer Security Applications
  Conference\/} (2015), ACM, pp.~101--110.

\bibitem{quach2017spiglass}
{\sc Quach, A., Cole, M., and Prakash, A.}
\newblock Supplementing modern software defenses with stack-pointer sanity.
\newblock In {\em Proceedings of the 33rd Annual Computer Security Applications
  Conference\/} (2017), ACSAC 2017.

\bibitem{quach2017study}
{\sc Quach, A., Erinfolami, R., Demicco, D., and Prakash, A.}
\newblock A multi-os cross-layer study of bloating in user programs, kernel and
  managed execution environments.
\newblock In {\em Proceedings of the 2017 Workshop on Forming an Ecosystem
  Around Software Transformation\/} (New York, NY, USA, 2017), FEAST '17, ACM,
  pp.~65--70.

\bibitem{salwan:rop}
{\sc Salwan, J.}
\newblock Ropgadget tool, 2012.
\newblock {\em URL http://shell-storm. org/project/ROPgadget\/}.

\bibitem{shacham:2007:turingrop}
{\sc Shacham, H.}
\newblock The geometry of innocent flesh on the bone: Return-into-libc without
  function calls (on the x86).
\newblock In {\em Proceedings of the 14th ACM conference on Computer and
  communications security\/} (2007), ACM, pp.~552--561.

\bibitem{sotirov2007heap}
{\sc Sotirov, A.}
\newblock Heap feng shui in javascript.

\bibitem{steensgaard1996points}
{\sc Steensgaard, B.}
\newblock Points-to analysis in almost linear time.
\newblock In {\em Proceedings of the 23rd ACM SIGPLAN-SIGACT symposium on
  Principles of programming languages\/} (1996), ACM, pp.~32--41.

\bibitem{sui2016svf}
{\sc Sui, Y., and Xue, J.}
\newblock Svf: Interprocedural static value-flow analysis in llvm.
\newblock In {\em Proceedings of the 25th International Conference on Compiler
  Construction\/} (New York, NY, USA, 2016), CC 2016, ACM, pp.~265--266.

\bibitem{team2003pax}
{\sc Team, P.}
\newblock Pax address space layout randomization (aslr).

\bibitem{vtvgcc:2012:tice}
{\sc Tice, C., Roeder, T., Collingbourne, P., Checkoway, S., Erlingsson,
  {\'U}., Lozano, L., and Pike, G.}
\newblock {Enforcing Forward-Edge Control-Flow Integrity in GCC \& LLVM}.
\newblock In {\em Proceedings of 23rd USENIX Security Symposium (USENIX
  Security'14)\/} (2014), pp.~941--955.

\bibitem{van2015practical}
{\sc van~der Veen, V., Andriesse, D., G{\"o}kta{\c{s}}, E., Gras, B., Sambuc,
  L., Slowinska, A., Bos, H., and Giuffrida, C.}
\newblock Practical context-sensitive cfi.
\newblock In {\em Proceedings of the 22nd ACM SIGSAC Conference on Computer and
  Communications Security\/} (2015), ACM, pp.~927--940.

\bibitem{wartell:2012:stir}
{\sc Wartell, R., Mohan, V., Hamlen, K.~W., and Lin, Z.}
\newblock Binary stirring: Self-randomizing instruction addresses of legacy x86
  binary code.
\newblock In {\em Proceedings of the 2012 ACM conference on Computer and
  communications security (CCS'12)\/} (2012), ACM, pp.~157--168.

\bibitem{whaley2004cspta}
{\sc Whaley, J., and Lam, M.~S.}
\newblock Cloning-based context-sensitive pointer alias analysis using binary
  decision diagrams.
\newblock {\em SIGPLAN Not. 39}, 6 (June 2004), 131--144.

\bibitem{wilson1995cspta}
{\sc Wilson, R.~P., and Lam, M.~S.}
\newblock Efficient context-sensitive pointer analysis for c programs.
\newblock In {\em Proceedings of the ACM SIGPLAN 1995 Conference on Programming
  Language Design and Implementation\/} (New York, NY, USA, 1995), PLDI '95,
  ACM, pp.~1--12.

\bibitem{xu2010software}
{\sc Xu, G., Mitchell, N., Arnold, M., Rountev, A., and Sevitsky, G.}
\newblock Software bloat analysis: finding, removing, and preventing
  performance problems in modern large-scale object-oriented applications.
\newblock In {\em Proceedings of the FSE/SDP workshop on Future of software
  engineering research\/} (2010), ACM, pp.~421--426.

\bibitem{xu2008cspta}
{\sc Xu, G., and Rountev, A.}
\newblock Merging equivalent contexts for scalable heap-cloning-based
  context-sensitive points-to analysis.
\newblock In {\em Proceedings of the 2008 International Symposium on Software
  Testing and Analysis\/} (New York, NY, USA, 2008), ISSTA '08, ACM,
  pp.~225--236.

\bibitem{yu2010fspta}
{\sc Yu, H., Xue, J., Huo, W., Feng, X., and Zhang, Z.}
\newblock Level by level: Making flow- and context-sensitive pointer analysis
  scalable for millions of lines of code.
\newblock In {\em Proceedings of the 8th Annual IEEE/ACM International
  Symposium on Code Generation and Optimization\/} (New York, NY, USA, 2010),
  CGO '10, ACM, pp.~218--229.

\bibitem{Zhang:2016:Vtrust}
{\sc Zhang, C., Carr, S.~A., Li, T., Ding, Y., Song, C., Payer, M., and Song,
  D.}
\newblock Vtrust: Regaining trust on virtual calls.
\newblock In {\em Symposium on Network and Distributed System Security
  (NDSS'16)\/} (2016).

\bibitem{Lenx:2013:CFI:Bin}
{\sc Zhang, C., Wei, T., Chen, Z., Duan, L., Szekeres, L., McCamant, S., Song,
  D., and Zou, W.}
\newblock Practical control flow integrity and randomization for binary
  executables.
\newblock In {\em Proceedings of the IEEE Symposium on Security and Privacy
  (Oakland'13)\/} (2013), pp.~559--573.

\bibitem{Zhang:2013:CFI:Bin}
{\sc Zhang, M., and Sekar, R.}
\newblock Control flow integrity for {COTS} binaries.
\newblock In {\em Proceedings of the 22nd USENIX Security Symposium (Usenix
  Security'13)\/} (2013), pp.~337--352.

\end{thebibliography}
\end{document}